\documentclass[twocolumn,prb]{revtex4}%
\usepackage{amsfonts}
\usepackage{amsmath}
\usepackage{amssymb}
\usepackage{graphicx}%
\providecommand{\U}[1]{\protect\rule{.1in}{.1in}}

\begin{document}
\preprint{ }
\title[All-coupling polaron optical response]{All-coupling polaron optical response: analytic approaches beyond the
adiabatic approximation}
\author{S. N. Klimin}
\altaffiliation{On leave from Department of Theoretical Physics, State University of Moldova,
MD-2009 Kishinev, Republic of Moldova.}

\author{J. Tempere}
\altaffiliation{Lyman Laboratory of Physics, Harvard University}

\author{J. T. Devreese}
\altaffiliation{Technische Universiteit Eindhoven, P.O. Box 513, 5600 MB Eindhoven, The Netherlands.}

\affiliation{TQC, Universiteit Antwerpen, Universiteitsplein 1, B-2610 Antwerpen, Belgium}


\begin{abstract}
In the present work, the problem of an all-coupling analytic description for
the optical conductivity of the Fr\"{o}hlich polaron is treated, with the goal
being to bridge the gap in validity range that exists between two
complementary methods: on the one hand the memory function formalism and on
the other hand the strong-coupling expansion based on the Franck-Condon
picture for the polaron response. At intermediate coupling, both methods were
found to fail as they do not reproduce Diagrammatic Quantum Monte Carlo
results. To resolve this, we modify the memory function formalism with respect
to the Feynman-Hellwarth-Iddings-Platzman (FHIP) approach, in order to take
into account a non-quadratic interaction in a model system for the polaron.
The strong-coupling expansion is extended beyond the adiabatic approximation,
by including into the treatment non-adiabatic transitions between excited
polaron states. The polaron optical conductivity that we obtain by combining
the two extended methods agree well, both qualitatively and quantitatively,
with the Diagrammatic Quantum Monte Carlo results in the whole available range
of the electron-phonon coupling strength.

\end{abstract}
\date{\today}
\maketitle

\section{Introduction \label{sec:Intro}}

The polaron, first proposed as a physical concept by L. D. Landau
\cite{Landau} in the context of electrons in polar crystals, has become a
generic notion describing a particle interacting with a quantized bosonic
field. The polaron problem has consequently been used for a long time as a
testing ground for various analytic and numerical methods with applications in
quantum statistical physics and quantum field theory. In condensed matter
physics, the polaron effect coming from the electron-phonon interaction is a
necessary ingredient in the description of the DC mobility and the optical
response in polar crystals (see Ref. \cite{Devreese2009}). Polaronic effects
are manifest in many interesting systems, such as magnetic polarons
\cite{Hemolt}, polarons in semiconducting polymers \cite{Sirringhaus}, and
complex oxides \cite{Franchini1,Franchini2015} which are described in terms of
the small-polaron theory \cite{Holstein}. \emph{Large-polaron} theory has
recently been stimulated by the possibility to study polaronic effects using
highly tunable quantum gases: the physics of an impurity immersed in an atomic
Bose-Einstein condensate\cite{BECpol2} can be modeled on the basis of a
Fr\"{o}hlich Hamiltonian. Another recent development in large-polaron physics
stems from the experimental advances in the determination of the band
structure of highly polar oxides \cite{Meevasana}, relevant for
superconductivity, where the optical response of complex oxides explicitly
shows the large-polaron features \cite{Mechelen2008,PRB2010}.

Diagrammatic Quantum Monte Carlo (DQMC) methods have been applied in recent
years to numerically calculate the ground state energy and the optical
conductivity of the Fr\"{o}hlich polaron \cite{M2000,M2003}. Advances in
computational techniques such as DQMC inspired renewed study of the key
problem in polaron theory -- an \emph{analytic} description of the polaron
response. For the \emph{small-polaron} optical conductivity, the all-coupling
analytic theory has been successfully developed \cite{Berciu} showing good
agreement with the numeric results of the DQMC. However, the optical response
problem for a \emph{large polaron} is not yet completely solved analytically.

Asymptotically exact analytic solutions for the polaron optical conductivity
have been obtained in the limits of weak \cite{GLF,DHL1971} and strong
coupling \cite{PRL2006,PRB2014}. A first proposal for an all-coupling
approximation for the polaron optical conductivity has been formulated in Ref.
\cite{DSG} (below referred to as DSG), further developing the
Feynman-Hellwarth-Iddings-Platzman theory \cite{FHIP} (FHIP) and using the
Feynman variational approach \cite{Feynman}. However, in Ref. \cite{DSG}, it
was already demonstrated that FHIP at large $\alpha$ is inconsistent with the
Heisenberg uncertainty relations. This inconsistency is revealed in Ref.
\cite{DSG} through extremely narrow peaks of the optical conductivity at large
$\alpha$. Nevertheless, the peak positions for the polaron optical
conductivity as obtained in Ref. \cite{DSG} have been confirmed with high
accuracy \cite{PRB2014} by the DQMC calculation \cite{M2003}. This inspired
further attempts to develop analytical methods for the polaron optical
response, especially at intermediate and strong coupling. Among these analytic
methods, an extension of the DSG method has been proposed in Ref.
\cite{PRL2006} introducing an extended memory function formalism with a
relaxation time determined from the additional sum rule for the polaron
optical conductivity. Alternatively, for the strong coupling regime, the
strong coupling expansion (SCE) based on the Franck-Condon scheme for
multiphonon optical conductivity has been developed in Refs.
\cite{PRL2006,PRB2014}.

The extended memory-function formalism and the strong coupling expansion of
Ref. \cite{PRL2006} are complementary to each other. The memory-function
formalism is well-substantiated for small and intermediate values of $\alpha$,
and the strong coupling expansion adequately describes the opposite limit of
large $\alpha$. However these two methods only qualitatively agree with each
other and with the DQMC data in the range of intermediate coupling strengths.
On one hand, the memory function formalism disagrees with DQMC at large
$\alpha$. On the other hand, the strong-coupling expansion only qualitatively
reproduces the shape of the optical conductivity and fails at intermediate
$\alpha$. The main aim of the current paper is to \emph{extend both the memory
function formalism and the strong coupling expansion in order to bridge the
gap that remains between their regions of validity}, such that the combination
of both methods allows to find analytical results in agreement with the
numeric DQMC results at all coupling.

We have added the following new elements in the theory which lead to an
overlapping of the areas of applicability for two aforesaid analytic methods.
For weak and intermediate coupling strengths, an extension of the Feynman
variational principle and the memory-function method for a polaron with a
non-quadratic trial action has been developed. As distinct from the memory
function formalism of Ref. \cite{PRL2006}, we do not use additional sum rules
and relaxation times, and perform the calculation \emph{ab initio}. For
intermediate and strong coupling strengths, the strong coupling expansion of
Ref. \cite{PRB2014} is extended beyond the adiabatic approximation accounting
for non-adiabatic transitions between excited polaron states. This leads to a
substantial expansion of the range of validity for the strong-coupling
expansion towards smaller $\alpha$ and to an overall improvement of its
agreement with DQMC.

The paper is organized as follows. In Sec. \ref{sec:Theory} we describe an
all-coupling analytic description for the polaron optical conductivity within
the extended memory function formalism with a non-parabolic trial action and
the non-adiabatic strong-coupling expansion. Sec. \ref{sec:Results} contains
the discussion of the obtained optical conductivity spectra and their
comparison with results of other methods and with the DQMC data. The
discussion is followed by conclusions, Sec. \ref{sec:Conclusions}.

\section{Analytic methods for the polaron optical conductivity
\label{sec:Theory}}

\subsection{Memory function formalism with a non-parabolic trial action}

To generalize the memory function formalism, we start by extending Feynman's
variational approach to translation invariant non-Gaussian trial actions. The
electron-phonon system is described by the Fr\"{o}hlich Hamiltonian, using the
Feynman units with $\hbar=1,$ the LO-phonon frequency $\omega_{\mathrm{LO}}%
=1$, and the band mass $m_{b}=1$,%
\begin{align}
\hat{H}  &  =\frac{\mathbf{\hat{p}}^{2}}{2}+\sum_{\mathbf{q}}\left(  \hat
{a}_{\mathbf{q}}^{+}\hat{a}_{\mathbf{q}}+\frac{1}{2}\right) \nonumber\\
&  +\frac{1}{\sqrt{V}}\sum_{\mathbf{q}}\frac{\sqrt{2\sqrt{2}\pi\alpha}}%
{q}\left(  \hat{a}_{\mathbf{q}}+\hat{a}_{-\mathbf{q}}^{+}\right)
e^{i\mathbf{q\cdot\hat{r}}}, \label{H}%
\end{align}
where $\mathbf{\hat{r}}$ is the position operator of the electron,
$\mathbf{\hat{p}}$ is its momentum operator; $\hat{a}_{\mathbf{q}}^{\dagger}$
and $\hat{a}_{\mathbf{q}}$ are, respectively, the creation and annihilation
operators for longitudinal optical (LO) phonons of wave vector $\mathbf{q}$.
The electron-phonon coupling strength is described by the Fr\"{o}hlich
coupling constant $\alpha$. As this Hamiltonian is quadratic in the phonon
degrees of freedom, they can be integrated out analytically in the
path-integral approach \cite{Feynman}. The remaining electron degree of
freedom is described via an action functional where the effects of
electron-phonon interaction are contained in an influence phase:%
\begin{equation}
S[\mathbf{r}_{e}(\tau)]=\frac{1}{2}%
{\displaystyle\int\limits_{0}^{\beta}}
\mathbf{\dot{r}}_{e}^{2}(\tau)d\tau-\Phi\lbrack\mathbf{r}_{e}(\tau)].
\label{Strue}%
\end{equation}
Here $\mathbf{r}_{e}(\tau)$ is the path of the electron, expressed in
imaginary time so as to obtain the euclidean action, and $\beta=1/(k_{B}T)$
with $T$ the temperature. The influence phase corresponding to (\ref{H}) is
\begin{align}
\Phi\lbrack\mathbf{r}_{e}(\tau)]  &  =\sqrt{2}\pi\alpha\int\frac{d\mathbf{q}%
}{(2\pi)^{3}}%
{\displaystyle\int\limits_{0}^{\beta}}
d\tau%
{\displaystyle\int\limits_{0}^{\beta}}
d\tau^{\prime}\frac{\cosh\left(  \left\vert \tau-\tau^{\prime}\right\vert
-\frac{\beta}{2}\right)  }{\sinh(\beta/2)}\nonumber\\
&  \times e^{i\mathbf{q}\cdot\left[  \mathbf{r}_{e}(\tau)-\mathbf{r}_{e}%
(\tau^{\prime})\right]  }. \label{Phi}%
\end{align}
This depends on the difference in electron position at different times,
resulting in a retarded action functional. In the path-integral formalism,
thermodynamic potentials (such as the free energy) are calculated via the
partition sum, which in turn is written as a sum over all possible paths
$\mathbf{r}_{e}(\tau)$ of the electron that start and end in the same point,
weighted by the exponent of the action:
\begin{equation}
e^{-\beta F}=\mathcal{Z}=\int\mathcal{D}\mathbf{r}_{e}e^{-S[\mathbf{r}%
_{e}(\tau)]}.
\end{equation}

\bigskip

Feynman's original variational method considers a \emph{quadratic} trial
action where the phonon degrees of freedom are replaced a fictitious particle
with coordinate $\mathbf{r}_{f}(\tau)$,
\begin{align}
S_{\text{quad}}\left[  \mathbf{r}_{e}(\tau),\mathbf{r}_{f}(\tau)\right]   &  =%
{\displaystyle\int\limits_{0}^{\beta}}
\left[  \frac{m\mathbf{\dot{r}}_{e}^{2}}{2}\right.  \nonumber\\
&  \left.  +\frac{m_{f}\mathbf{\dot{r}}_{f}^{2}}{2}+V\left(  \mathbf{r}%
_{f}-\mathbf{r}_{e}\right)  \right]  d\tau,
\end{align}
interacting with the electron through a harmonic potential:%
\begin{equation}
V\left(  \mathbf{r}_{f}-\mathbf{r}_{e}\right)  =\frac{m_{f}\omega^{2}}%
{2}\left(  \mathbf{r}_{f}-\mathbf{r}_{e}\right)  ^{2}.
\end{equation}
The partition sum corresponding to this model is
\begin{equation}
\mathcal{Z}_{\text{quad}}=\int\mathcal{D}\mathbf{r}_{e}\int\mathcal{D}%
\mathbf{r}_{f}e^{-S_{\text{quad}}[\mathbf{r}_{e}(\tau),\mathbf{r}_{f}(\tau)]}.
\end{equation}
Expectation values of of functionals $\mathcal{F}[\mathbf{r}_{e}(\tau)]$\ of
the electron path are given by%
\begin{align}
\left\langle \mathcal{F}[\mathbf{r}_{e}(\tau)]\right\rangle _{\text{quad}} &
=\frac{1}{\mathcal{Z}_{\text{quad}}}\int\mathcal{D}\mathbf{r}_{e}\text{
}\mathcal{F}[\mathbf{r}_{e}(\tau)]\nonumber\\
&  \times\int\mathcal{D}\mathbf{r}_{f}e^{-S_{\text{quad}}[\mathbf{r}_{e}%
(\tau),\mathbf{r}_{f}(\tau)]}.
\end{align}
In the above formula, it is clear that performing the path integral over
$\mathbf{r}_{f}$ exactly may simplify the result. The result of this
integration still depends on the path $\mathbf{r}_{e}(\tau)$ and is written in
terms of an influence phase,%
\begin{align}
&  \int\mathcal{D}\mathbf{r}_{f}\exp\left\{  -%
{\displaystyle\int\limits_{0}^{\beta}}
\left[  \frac{m_{f}\mathbf{\dot{r}}_{f}^{2}}{2}+\frac{m_{f}\omega^{2}}%
{2}\left(  \mathbf{r}_{f}-\mathbf{r}_{e}\right)  ^{2}\right]  d\tau\right\}
\nonumber\\
&  =\mathcal{Z}_{f}\exp\left\{  \Phi_{\text{quad}}[\mathbf{r}_{e}%
(\tau)]\right\}  ,\label{fictint}%
\end{align}
where $\mathcal{Z}_{f}$ is a constant independent of $\mathbf{r}_{e}(\tau)$.
The influence phase corresponds to a quadratic, retarded interaction for the
electron. The model system partition sum becomes
\begin{equation}
\mathcal{Z}_{\text{quad}}=\mathcal{Z}_{f}\int\mathcal{D}\mathbf{r}_{e}%
\exp\left\{  -%
{\displaystyle\int\limits_{0}^{\beta}}
\frac{m\mathbf{\dot{r}}_{e}^{2}}{2}d\tau+\Phi_{\text{quad}}[\mathbf{r}%
_{e}(\tau)]\right\}  .
\end{equation}
Feynman restricted his trial action to a quadratic action, since only for case
one can calculate the influence phase analytically, and obtain
\begin{align}
\Phi_{\text{quad}}[\mathbf{r}_{e}(\tau)] &  =-\frac{m_{f}\omega^{3}}{4}%
{\displaystyle\int\limits_{0}^{\beta}}
d\tau%
{\displaystyle\int\limits_{0}^{\beta}}
d\tau^{\prime}\left[  \mathbf{r}_{e}(\tau)-\mathbf{r}_{e}(\tau^{\prime
})\right]  ^{2}\nonumber\\
&  \times\frac{\cosh\left[  \omega\left(  \left\vert \tau-\tau^{\prime
}\right\vert -\frac{\beta}{2}\right)  \right]  }{\sinh(\beta\omega/2)}.
\end{align}

The essence of the Feynman variational method consists in writing the
partition function of the true electron-phonon system (\ref{Strue}) as%
\begin{align}
\mathcal{Z}  &  =\frac{1}{\mathcal{Z}_{f}}\int\mathcal{D}\mathbf{r}_{e}%
\exp\left\{  \Phi\lbrack\mathbf{r}_{e}(\tau)]-\Phi_{\text{quad}}%
[\mathbf{r}_{e}(\tau)]\right\} \nonumber\\
&  \times\int\mathcal{D}\mathbf{r}_{f}\exp\left\{  -S_{\text{quad}}%
[\mathbf{r}_{e}(\tau),\mathbf{r}_{f}(\tau)]\right\}  . \label{moot}%
\end{align}
Indeed, performing the path integration for the fictitious particle via
(\ref{fictint}) cancels $\Phi_{\text{quad}}[\mathbf{r}_{e}(\tau)]$ as well as
the factor $\mathcal{Z}_{f}$, and leaves the kinetic energy contribution,
restoring the action function of the true system. The usefulness of the above
expression lies in the fact that it can also be interpreted as an expectation
value with respect to the model system:
\begin{equation}
\mathcal{Z}=\frac{\mathcal{Z}_{\text{quad}}}{\mathcal{Z}_{f}}\left\langle
\exp\left\{  \Phi\lbrack\mathbf{r}_{e}(\tau)]-\Phi_{\text{quad}}%
[\mathbf{r}_{e}(\tau)]\right\}  \right\rangle _{\text{quad}}.
\end{equation}
Now using Jensen's inequality
\begin{equation}
\left\langle e^{\mathcal{F}[\mathbf{r}_{e}(\tau)]}\right\rangle \geqslant
e^{\left\langle \mathcal{F}[\mathbf{r}_{e}(\tau)]\right\rangle },
\end{equation}
and taking the logarithm of the resulting expression leads to
\begin{equation}
F\leqslant F_{0}+\frac{1}{\beta}\left\langle \Phi_{\text{quad}}[\mathbf{r}%
_{e}(\tau)]-\Phi\lbrack\mathbf{r}_{e}(\tau)]\right\rangle _{\text{quad}},
\label{JensFeyn}%
\end{equation}
which is the Jensen-Feynman variational inequality. Here we introduce the
notation $\mathcal{Z}_{\text{quad}}/\mathcal{Z}_{f}=e^{-\beta F_{0}}$. Note
that when we write the above expression in terms of the actions rather than
the influence phases, one gets the more familiar form of the Jensen-Feynman
inequality:%
\begin{equation}
F\leqslant F_{0}+\frac{1}{\beta}\left\langle S[\mathbf{r}_{e}(\tau
)]-S_{\text{quad}}[\mathbf{r}_{e}(\tau)]\right\rangle _{\text{quad}}.
\end{equation}
Using the Feynman variational approach with the Gaussian trial action,
excellent results are obtained for the polaron ground-state energy, free
energy, and the effective mass. Moreover, this approach has been effectively
used to derive the DSG\ all-coupling theory for the polaron optical
conductivity, Ref. \cite{DSG}. However, as mentioned in the introduction, the
DSG and DQMC results contradict to each other in the range of large $\alpha$.
The most probable source of this contradiction is the Gaussian form of the
trial action used in the DSG theory. Indeed, the model system contains only a
single frequency, leading to unphysically sharp peaks in the spectrum, subject
to thermal broadening only \cite{Dries1,Dries2}. Extensions to the formalism
\cite{PRL2006} have tried to overcome this problem by including an ad-hoc
broadening of the energy level, chosen in such as way as to comply with the
sum rules. A remarkable success in the problem of the polaron optical response
has been achieved in the recent work \cite{DS}, where the all-coupling polaron
optical conductivity is calculated using the general quadratic trial action
instead of the Feynman model with a single fictitious particle. The resulting
optical conductivity is in good agreement with DQMC results \cite{M2003} in
the weak- and intermediate-coupling regimes and is qualitatively in line with
DQMC even at extremely strong coupling, resolving the issue of the linewidth
in the FHIP approach.

In the literature, there are attempts to re-formulate the Feynman variational
approach avoiding retarded trial actions. For example, Cataudella et al.
\cite{Catau} introduce an extended action which contains the coordinates of
the electron, the fictitious particle, and the phonons. This action, however,
is not exactly equivalent to the action of the electron-phonon system, and
hence the results obtained in \cite{Catau} need verification. In Ref.
\cite{SSC}, we introduced an extended action/Hamiltonian for an
electron-phonon system and reformulated the Feynman variational method in the
Hamiltonian representation. This method leads to the same result as the
Feynman variational approach. However the method of Ref. \cite{SSC} reproduces
the strong coupling limit for the polaron energy only when using a Gaussian
trial action.

In the current work, we propose to extend the Feynman variational approach to
trial systems with non-parabolic interactions between an electron and a
fictitious particle. The difficulty with using non-Gaussian trial actions is
that the influence phase (here $\Phi_{\text{quad}}$) can only be computed
analytically for quadratic action functionals. However, quantum-statistical
expectation values (such as the one in the Jensen-Feynman inequality) can be
calculated for non-quadratic model systems by other means, in particular if
the spectrum of eigenvalues and eigenfunctions can be found.\textbf{ }So, what
we propose is \emph{to focus on keeping the influence phase for a quadratic
model system in the expressions, while at the same time allowing for
non-Gaussian potentials for the expectation values}.

Consider a (non-quadratic) variational trial action
\begin{align}
S_{\text{var}}\left[  \mathbf{r}_{e}(\tau),\mathbf{r}_{f}(\tau)\right]   &  =%
{\displaystyle\int\limits_{0}^{\beta}}
\left[  \frac{m\mathbf{\dot{r}}_{e}^{2}}{2}\right.  \nonumber\\
&  \left.  +\frac{m_{f}\mathbf{\dot{r}}_{f}^{2}}{2}+U\left(  \mathbf{r}%
_{f}-\mathbf{r}_{e}\right)  \right]  d\tau
\end{align}
with a general potential $U$. Since
\begin{align}
S_{\text{quad}}\left[  \mathbf{r}_{e}(\tau),\mathbf{r}_{f}(\tau)\right]   &
=S_{\text{var}}\left[  \mathbf{r}_{e}(\tau),\mathbf{r}_{f}(\tau)\right]
\nonumber\\
&  +%
{\displaystyle\int\limits_{0}^{\beta}}
\left[  V\left(  \mathbf{r}_{f}-\mathbf{r}_{e}\right)  -U\left(
\mathbf{r}_{f}-\mathbf{r}_{e}\right)  \right]  d\tau,
\end{align}
\ we can rewrite (\ref{moot}) to:%
\begin{align}
\mathcal{Z} &  =\frac{1}{\mathcal{Z}_{f}}\int\mathcal{D}\mathbf{r}_{e}%
\int\mathcal{D}\mathbf{r}_{f}\exp\left\{  \Phi\lbrack\mathbf{r}_{e}%
(\tau)]-\Phi_{\text{quad}}[\mathbf{r}_{e}(\tau)]\right\}  \nonumber\\
&  \times\exp\left\{  -%
{\displaystyle\int\limits_{0}^{\beta}}
\left[  V\left(  \mathbf{r}_{f}-\mathbf{r}_{e}\right)  -U\left(
\mathbf{r}_{f}-\mathbf{r}_{e}\right)  \right]  d\tau\right\}  \nonumber\\
&  \times\exp\left\{  -S_{\text{var}}\left[  \mathbf{r}_{e}(\tau
),\mathbf{r}_{f}(\tau)\right]  \right\}  .\label{tiki}%
\end{align}
Expectation values of functionals of $\mathbf{r}_{e}(\tau)$ and $\mathbf{r}%
_{f}(\tau)$ with respect to the non-quadratic variational model system are
given by%
\begin{align}
\left\langle \mathcal{F}[\mathbf{r}_{e},\mathbf{r}_{f}]\right\rangle
_{\text{var}} &  =\frac{1}{\mathcal{Z}_{\text{var}}}\int\mathcal{D}%
\mathbf{r}_{e}\int\mathcal{D}\mathbf{r}_{f}\mathcal{F}[\mathbf{r}%
_{e},\mathbf{r}_{f}]\nonumber\\
&  \times\exp\left\{  -S_{\text{var}}\left[  \mathbf{r}_{e}(\tau
),\mathbf{r}_{f}(\tau)\right]  \right\}  .\label{expval}%
\end{align}
This allows to interpret expression (\ref{tiki}) as
\begin{align}
\mathcal{Z} &  =\frac{\mathcal{Z}_{\text{var}}}{\mathcal{Z}_{f}}\left\langle
\exp\left\{  \Phi\lbrack\mathbf{r}_{e}(\tau)]-\Phi_{\text{quad}}%
[\mathbf{r}_{e}(\tau)]\right.  \right.  \nonumber\\
&  \left.  \left.  -%
{\textstyle\int\nolimits_{0}^{\beta}}
\left[  V\left(  \mathbf{r}_{f}-\mathbf{r}_{e}\right)  -U\left(
\mathbf{r}_{f}-\mathbf{r}_{e}\right)  \right]  d\tau\right\}  \right\rangle
_{\text{var}}%
\end{align}
With $\mathcal{Z}_{\text{var}}/\mathcal{Z}_{f}=e^{-\beta F_{\text{var}}}$ and
using again Jensen's inequality, we arrive at:%
\begin{align}
F &  \leqslant F_{\text{var}}+\frac{1}{\beta}\left\langle \Phi_{\text{quad}%
}[\mathbf{r}_{e}(\tau)]-\Phi\lbrack\mathbf{r}_{e}(\tau)]\right\rangle
_{\text{var}}\nonumber\\
&  +\left\langle V\left(  \mathbf{r}_{f}-\mathbf{r}_{e}\right)  -U\left(
\mathbf{r}_{f}-\mathbf{r}_{e}\right)  \right\rangle _{\text{var}}\label{ineq2}%
\end{align}
We have used that for time-independent potentials, $\frac{1}{\beta}%
{\textstyle\int\nolimits_{0}^{\beta}}
d\tau=1$. When $U=V$, and we choose a quadratic interaction potential, this
restores the original Jensen-Feynman variational principle. However,
deviations from a quadratic potential result in two changes. Firstly, there is
an additional term corresponding to the expectation value of the difference
between the chosen variational potential and the quadratic one. Secondly, the
expectation values are to be calculated with respect to the chosen variational
potential $U$ rather than with respect to the quadratic potential. We again
emphasize that the calculation of such expectation values does not require a
quadratic potential. \ Thus the \emph{new variational inequality}
(\ref{ineq2}) is an extension of the Feynman -- Jensen inequality.

It is important for the calculations that $S_{\text{var}}$ is translation
invariant but non-retarded action, so that all expressions in the variational
functional (\ref{ineq2}) have the same form in both representations -- path
integral and standard quantum mechanics. Apart from the parameters appearing
in the trial action $S_{\text{var}}$, the inequality (\ref{ineq2}) still
contains as variational parameters $m_{f}$ and $\omega$, inherited from the
\textquotedblleft auxiliary\textquotedblright\ quadratic action
$S_{\text{quad}}$ and appearing in $\Phi_{\text{quad}}$ and $V\left(
\mathbf{r}_{f}-\mathbf{r}_{e}\right)  $. These do not depend on the parameters
of the electron-phonon interaction at all, and therefore the minimization of
the free energy with respect to $m_{f}$ and $\omega$ can be performed
\emph{before} the minimization of the whole variational functional with
respect to the remaining parameters of the non-Gaussian trial action
$S_{\text{var}}$.

The extended Jensen-Feynman inequality (\ref{ineq2}), despite having more
variational parameters, does not lead in general to a substantially lower
polaron free energy than the original Feynman result, except in the extremely
strong coupling regime, where the present variational functional analytically
tends (for $T=0$) to the exact strong coupling limit obtained by Miyake
\cite{M75}. However, its advantage with respect to the original Feynman
treatment is in calculating the optical conductivity. A physically reasonable
choice of the trial interaction potential $U\left(  \mathbf{r}_{f}%
-\mathbf{r}\right)  $ is no longer restricting to a single frequency
oscillator. According to Refs. \cite{M75,Pekar1954}, the self-consistent
potential for an electron induced by the lattice polarization is parabolic
near the bottom and Coulomb-like at large distances. Therefore, for the
calculation of the optical conductivity, we choose a trial potential in the
piecewise form, stitching together a parabolic and a Coulomb-like potential in
a continuous way. The spectrum of internal states of the model system with
this potential necessarily consists of an infinite number non-equidistant
energy levels with the energies $E_{n}<0$ (counted from the potential energy
at the infinity distance from the polaron) and a continuum of energies $E>0$.
Accounting for transitions between all these levels, one must expect a
significant broadening of the peak absorption.

The polaron optical conductivity is calculated using the memory-function
formalism which differs from that applied in Refs. \cite{DSG,PD1983,PRL2006}
by using the non-quadratic trial action described above. The optical
conductivity of a gas of interacting polarons within the memory-function
technique, is given by the formula which is structurally similar to the
polaron optical conductivity \cite{DSG},%
\begin{equation}
\sigma\left(  \Omega\right)  =\frac{e^{2}n_{0}}{m_{b}}\frac{i}{\Omega
-\chi\left(  \Omega\right)  /\Omega}, \label{4}%
\end{equation}
where $n_{0}=N/V$ is the carrier density. The memory function in the
non-quadratic setting is given by%
\begin{align}
\chi\left(  \Omega\right)   &  =\frac{2}{3\hbar m_{b}}\int\frac{d\mathbf{q}%
}{\left(  2\pi\right)  ^{3}}q^{2}\left\vert V_{\mathbf{q}}\right\vert ^{2}%
\int_{0}^{\infty}dt\left(  e^{i\Omega t}-1\right) \nonumber\\
&  \times\operatorname{Im}\left[  \frac{\cos\left[  \omega_{0}\left(
t+\frac{i\hbar\beta}{2}\right)  \right]  }{\sinh\left(  \frac{\beta\hbar
\omega_{0}}{2}\right)  }\left\langle e^{i\mathbf{q\cdot r}\left(  t\right)
}e^{-i\mathbf{q\cdot r}}\right\rangle _{\text{var}}\right]  , \label{xi}%
\end{align}
where $\omega_{0}$ is the LO phonon frequency, and the correlation function
$\left\langle e^{i\mathbf{q\cdot r}\left(  t\right)  }e^{-i\mathbf{q\cdot r}%
}\right\rangle _{\text{var}}$ is calculated with the quantum states of the
trial Hamiltonian corresponding to $S_{\text{var}}$. In the particular case
$T=0$ we apply the formula following from (\ref{xi})%
\begin{align}
\chi\left(  \Omega\right)   &  =\frac{1}{3\pi^{2}\hbar m_{b}}\lim
_{\delta\rightarrow0_{+}}\int_{0}^{\infty}dq~\left\vert V_{\mathbf{q}%
}\right\vert ^{2}q^{4}\int\limits_{0}^{\infty}dt~e^{-\delta t}\nonumber\\
&  \times\left(  e^{i\Omega t}-1\right)  \operatorname{Im}\left(
e^{-i\omega_{0}t}\left\langle e^{i\mathbf{q\cdot r}\left(  t\right)
}e^{-i\mathbf{q\cdot r}}\right\rangle _{\text{var}}\right)  . \label{xia}%
\end{align}
Rather than computing the correlation function $\left\langle
e^{i\mathbf{q\cdot r}\left(  t\right)  }e^{-i\mathbf{q\cdot r}}\right\rangle
_{\text{var}}$ as a path integral through (\ref{expval}), we choose to
evaluate it in the equivalent Hamiltonian formalism. In this Hamiltonian
framework, (\ref{xia}) is written as a sum over the eigenstates of the trial
Hamiltonian for the electron and the fictitious particle interacting through
the potential $U$,%
\begin{equation}
\hat{H}_{\text{var}}=\frac{\mathbf{\hat{p}}^{2}}{2}+\frac{\mathbf{\hat{p}}%
_{f}^{2}}{2m_{f}}+U\left(  \mathbf{\hat{r}}_{f}-\mathbf{\hat{r}}\right)  .
\label{Hu}%
\end{equation}
Thus the correlation function is given by:%
\begin{align}
\left\langle e^{i\mathbf{q\cdot r}\left(  t\right)  }e^{-i\mathbf{q\cdot r}%
}\right\rangle _{\text{var}}  &  =\sum_{\mathbf{k}^{\prime};l^{\prime
},n^{\prime},m^{\prime}}e^{i\frac{t}{\hbar}\left[  \left(  \varepsilon
_{0,0}-\varepsilon_{l^{\prime},n^{\prime}}\right)  -\frac{\hbar^{2}\left(
\mathbf{k}^{\prime}\right)  ^{2}}{2M}\right]  }\nonumber\\
&  \times\left\vert \left\langle \psi_{\mathbf{0};0,0,0}\left\vert
e^{i\mathbf{q\cdot r}}\right\vert \psi_{\mathbf{k}^{\prime};l^{\prime
},n^{\prime},m^{\prime}}\right\rangle \right\vert ^{2}, \label{corfun}%
\end{align}
where $M=1+m_{f}$ is the total mass of the trial system and $\left\vert
\psi_{\mathbf{k};l,n,m}\right\rangle $ are the wave functions of the trial
system. The wave function $\left\vert \psi_{\mathbf{k};l,n,m}\right\rangle $
is factorized as a product of a plane wave for the center-of-mass motion (with
center-of-mass coordinate $\mathbf{R}$) and a wave function for the relative
motion $\left\vert \varphi_{l,n,m}\right\rangle $ (with the coordinate vector
$\boldsymbol{\rho}$ of the relative motion),%
\begin{align}
\left\vert \psi_{\mathbf{k};l,n,m}\right\rangle  &  =\frac{1}{\sqrt{V}%
}e^{i\mathbf{k\cdot R}}\left\vert \varphi_{l,n,m}\right\rangle ,\label{WF}\\
\left\vert \varphi_{l,n,m}\right\rangle  &  =\mathcal{R}_{l,n}\left(
\rho\right)  Y_{l,m}\left(  \theta,\varphi\right)  .
\end{align}
The quantum numbers for the Hamiltonian $\hat{H}_{\text{var}}$ are the
momentum $\mathbf{k}$, the quanta $l,m$ related to to angular momentum, and a
nodal quantum number $n$ for the relative motion wavefunction. The quantum
numbers $l,n$ determine the energy $\varepsilon_{l,n}$ associated with the
relative motion between electron and fictitious particle (including both the
discrete and continuous parts of the energy spectrum). When substituting
(\ref{corfun}) into the memory function we arrive at the result%
\begin{align}
\chi\left(  \Omega\right)   &  =\frac{1}{3\pi^{2}\hbar m_{b}}\int_{0}^{\infty
}dq~\left\vert V_{\mathbf{q}}\right\vert ^{2}q^{4}\int\limits_{0}^{\infty
}dte^{-\delta t}\nonumber\\
&  \times\sum_{\mathbf{k}^{\prime};l^{\prime},n^{\prime},m^{\prime}}\left\vert
\left\langle \psi_{\mathbf{0};0,0,0}\left\vert e^{i\mathbf{q\cdot r}%
}\right\vert \psi_{\mathbf{k}^{\prime};l^{\prime},n^{\prime},m^{\prime}%
}\right\rangle \right\vert ^{2}\nonumber\\
&  \times\left(  e^{i\Omega t}-1\right)  \operatorname{Im}\left(
e^{-i\frac{t}{\hbar}\left[  \left(  \varepsilon_{l^{\prime},n^{\prime}%
}-\varepsilon_{0,0}\right)  +\frac{\hbar^{2}\left(  \mathbf{k}^{\prime
}\right)  ^{2}}{2M}+\hbar\omega_{0}\right]  }\right)  , \label{xi1a}%
\end{align}
with $\delta\rightarrow+0$. Further, the Feynman units are used, where
$\hbar=1$, $\omega_{0}=1$, and the band mass $m_{b}=1$. In these units, the
squared modulus $\left\vert V_{\mathbf{q}}\right\vert ^{2}$ is%
\[
\left\vert V_{\mathbf{q}}\right\vert ^{2}=\frac{2\sqrt{2}\pi\alpha}{q^{2}}%
\]
The memory function is then given by the formula%
\begin{align}
\chi\left(  \Omega\right)   &  =\frac{2\sqrt{2}\alpha}{3\pi}\int_{0}^{\infty
}dq~q^{2}\nonumber\\
&  \times\sum_{\mathbf{k}^{\prime};l^{\prime},n^{\prime},m^{\prime}}\left\vert
\left\langle \psi_{\mathbf{0};0,0,0}\left\vert e^{i\mathbf{q\cdot r}%
}\right\vert \psi_{\mathbf{k}^{\prime};l^{\prime},n^{\prime},m^{\prime}%
}\right\rangle \right\vert ^{2}\nonumber\\
&  \times\int\limits_{0}^{\infty}dte^{-\delta t}\left(  e^{i\Omega t}-1\right)
\nonumber\\
&  \times\operatorname{Im}\left(  e^{-it\left[  \left(  \varepsilon
_{l^{\prime},n^{\prime}}-\varepsilon_{0,0}\right)  +\frac{1}{2M}\left(
\mathbf{k}^{\prime}\right)  ^{2}+1\right]  }\right)  . \label{xi4}%
\end{align}
The matrix element in (\ref{xi4}) is a product of two matrix elements:%
\begin{align}
&  \left\langle \psi_{\mathbf{k};l,n,m}\left\vert e^{i\mathbf{q\cdot r}%
}\right\vert \psi_{\mathbf{k}^{\prime};l^{\prime},n^{\prime},m^{\prime}%
}\right\rangle \nonumber\\
&  =\frac{1}{V}\left\langle e^{-i\mathbf{kR}}\left\vert e^{i\mathbf{q\cdot R}%
}\right\vert e^{i\mathbf{k}^{\prime}\mathbf{R}}\right\rangle \left\langle
\varphi_{l,n,m}\left\vert e^{i\mu\mathbf{q\cdot}\boldsymbol{\rho}}\right\vert
\varphi_{l^{\prime},n^{\prime},m^{\prime}}\right\rangle ,
\end{align}
where $\mu$ is the reduced mass of the trial system. The first matrix element
is%
\begin{equation}
\frac{1}{V}\left\langle e^{-i\mathbf{kR}}\left\vert e^{i\mathbf{q\cdot R}%
}\right\vert e^{i\mathbf{k}^{\prime}\mathbf{R}}\right\rangle =\delta
_{\mathbf{k}^{\prime},\mathbf{k}-\mathbf{q}}.
\end{equation}
This eliminates the integration over the final electron momentum
$\mathbf{k}^{\prime}$ and reduces the memory function to the expression%
\begin{align}
\chi\left(  \Omega\right)   &  =\frac{2\sqrt{2}\alpha}{3\pi}\int_{0}^{\infty
}dq~q^{2}\int\limits_{0}^{\infty}dte^{-\delta t}\nonumber\\
&  \times\sum_{l^{\prime},n^{\prime},m^{\prime}}\left\vert \left\langle
\varphi_{0,0,0}\left\vert e^{i\mu\mathbf{q\cdot}\boldsymbol{\rho}}\right\vert
\varphi_{l^{\prime},n^{\prime},m^{\prime}}\right\rangle \right\vert
^{2}\nonumber\\
&  \times\left(  e^{i\Omega t}-1\right)  \operatorname{Im}\left(
e^{-it\left(  \frac{q^{2}}{2M}+\varepsilon_{l^{\prime},n^{\prime}}%
-\varepsilon_{0,0}+1\right)  }\right)  .
\end{align}
The summation over $m^{\prime}$ is performed explicitly:%
\begin{align}
&  \sum_{m,m^{\prime}}\left\vert \left\langle \varphi_{l,n,m}\left\vert
e^{i\mu\mathbf{q\cdot}\boldsymbol{\rho}}\right\vert \varphi_{l^{\prime
},n^{\prime},m^{\prime}}\right\rangle \right\vert ^{2}\nonumber\\
&  =\frac{\left(  2l+1\right)  \left(  2l^{\prime}+1\right)  }{2}\int
_{0}^{\infty}\rho^{2}d\rho\int_{0}^{\infty}d\rho^{\prime}\left(  \rho^{\prime
}\right)  ^{2}\nonumber\\
&  \times\mathcal{R}_{l,n}\left(  \rho\right)  \mathcal{R}_{l^{\prime
},n^{\prime}}\left(  \rho\right)  \mathcal{R}_{l,n}\left(  \rho^{\prime
}\right)  \mathcal{R}_{l^{\prime},n^{\prime}}\left(  \rho^{\prime}\right)
\nonumber\\
&  \times\int_{0}^{2\pi}\frac{\sin\left(  \mu q\left\vert \boldsymbol{\rho
}-\boldsymbol{\rho}^{\prime}\right\vert \right)  }{\mu q\left\vert
\boldsymbol{\rho}-\boldsymbol{\rho}^{\prime}\right\vert }\nonumber\\
&  \times P_{l}\left(  \cos\theta\right)  P_{l^{\prime}}\left(  \cos
\theta\right)  \sin\theta d\theta. \label{e1}%
\end{align}
The modulus $\left\vert \boldsymbol{\rho}-\boldsymbol{\rho}^{\prime
}\right\vert $ is expressed as%
\begin{equation}
\left\vert \boldsymbol{\rho}-\boldsymbol{\rho}^{\prime}\right\vert =\sqrt
{\rho^{2}+\left(  \rho^{\prime}\right)  ^{2}-2\rho\rho^{\prime}\cos\theta}.
\end{equation}
Hence we can use the expansion of $\frac{\sin\left(  \mu q\left\vert
\boldsymbol{\rho}-\boldsymbol{\rho}^{\prime}\right\vert \right)  }{\mu
q\left\vert \boldsymbol{\rho}-\boldsymbol{\rho}^{\prime}\right\vert }$ through
the Legendre polynomials:%
\begin{align}
\frac{\sin\left(  \mu q\left\vert \boldsymbol{\rho}-\boldsymbol{\rho}^{\prime
}\right\vert \right)  }{\mu q\left\vert \boldsymbol{\rho}-\boldsymbol{\rho
}^{\prime}\right\vert }  &  =\sum_{l^{\prime\prime}=0}^{\infty}\left(
2l^{\prime\prime}+1\right) \nonumber\\
&  \times j_{l^{\prime\prime}}\left(  \mu q\rho\right)  j_{l^{\prime\prime}%
}\left(  \mu q\rho^{\prime}\right)  P_{l^{\prime\prime}}\left(  \cos
\theta\right)  .
\end{align}
The integral of the product of three Legendre polynomials is expressed through
the $3j$-symbol:
\begin{align}
&  \int_{0}^{2\pi}P_{l^{\prime\prime}}\left(  \cos\theta\right)  P_{l}\left(
\cos\theta\right)  P_{l^{\prime}}\left(  \cos\theta\right)  \sin\theta
d\theta\nonumber\\
&  =2\left(
\begin{array}
[c]{ccc}%
l & l^{\prime} & l^{\prime\prime}\\
0 & 0 & 0
\end{array}
\right)  ^{2}.
\end{align}
Therefore we find that%
\begin{align}
&  \sum_{m,m^{\prime}}\left\vert \left\langle \varphi_{l,n,m}\left\vert
e^{i\mu\mathbf{q\cdot}\boldsymbol{\rho}}\right\vert \varphi_{l^{\prime
},n^{\prime},m^{\prime}}\right\rangle \right\vert ^{2}\nonumber\\
&  =\sum_{l^{\prime\prime}=0}^{\infty}\left(  2l+1\right)  \left(  2l^{\prime
}+1\right)  \left(  2l^{\prime\prime}+1\right) \nonumber\\
&  \times\left(
\begin{array}
[c]{ccc}%
l & l^{\prime} & l^{\prime\prime}\\
0 & 0 & 0
\end{array}
\right)  ^{2}S_{q}^{2}\left(  l,n\left\vert l^{\prime\prime}\right\vert
l^{\prime},n^{\prime}\right)  ,
\end{align}
where $S_{q}\left(  l,n\left\vert l^{\prime\prime}\right\vert l^{\prime
},n^{\prime}\right)  $ is the matrix element with radial wave functions for
the trial system,%
\begin{align}
&  S_{q}\left(  l,n\left\vert l^{\prime\prime}\right\vert l^{\prime}%
,n^{\prime}\right) \nonumber\\
&  \equiv\int_{0}^{\infty}\mathcal{R}_{l,n}\left(  \rho\right)  \mathcal{R}%
_{l^{\prime},n^{\prime}}\left(  \rho\right)  j_{l^{\prime\prime}}\left(  \mu
q\rho\right)  \rho^{2}d\rho. \label{mel}%
\end{align}
For $l=0$ the result of the summation over intermediate states is reduced to
the formula%
\begin{align}
&  \sum_{m^{\prime}}\left\vert \left\langle \varphi_{0,n,0}\left\vert
e^{i\mu\mathbf{q\cdot}\boldsymbol{\rho}}\right\vert \varphi_{l^{\prime
},n^{\prime},m^{\prime}}\right\rangle \right\vert ^{2}\nonumber\\
&  =\left(  2l^{\prime}+1\right)  S_{q}^{2}\left(  0,0\left\vert l^{\prime
}\right\vert l^{\prime},n^{\prime}\right)  .
\end{align}
After this summation and the integration over time, the memory function takes
the form%
\begin{align}
\chi\left(  \Omega\right)   &  =\frac{\sqrt{2}\alpha}{3\pi}\int_{0}^{\infty
}dq~q^{2}\sum_{l,n}\left(  2l+1\right)  S_{q}^{2}\left(  0,0\left\vert
l\right\vert l,n\right) \nonumber\\
&  \times\left(  \frac{1}{\Omega-\Omega_{q,l,n}+i\delta}\right. \nonumber\\
&  \left.  -\frac{1}{\Omega+\Omega_{q,l,n}+i\delta}+\frac{2}{\Omega_{q,l,n}%
}\right)  .\label{xi5}\\
&  \left(  \delta\rightarrow+0\right) \nonumber
\end{align}
with the transition frequency for transitions between the ground and excited
states of the trial system accompanied by an emission of a phonon:%
\begin{equation}
\Omega_{q,l,n}\equiv\frac{q^{2}}{2M}+\varepsilon_{l,n}-\varepsilon_{0,0}+1.
\label{Wln}%
\end{equation}
Expression (\ref{xi5}) is used for the numerical calculation of the polaron
optical conductivity within the extended memory function formalism.

\subsection{Non-adiabatic strong coupling expansion}

Next, we describe the strong coupling approach and its extension beyond the
adiabatic approximation, denoted below as the \emph{non-adiabatic SCE}. Here,
the goal is to take non-adiabatic transitions between different excited levels
of a polaron into account in the formalism. The notations in this subsection
are the same as in Ref. \cite{PRB2014}. The polaron optical conductivity in
the strong coupling regime is represented by the Kubo formula,%
\begin{equation}
\operatorname{Re}\sigma\left(  \Omega\right)  =\frac{\Omega}{2}\int_{-\infty
}^{\infty}e^{i\Omega t}f_{zz}\left(  t\right)  \,dt, \label{KuboDD0}%
\end{equation}
with the dipole-dipole correlation function%
\begin{align}
f_{zz}\left(  t\right)   &  =\sum_{n,l,m,}\sum_{n^{\prime},l^{\prime
},m^{\prime},}\sum_{n^{\prime\prime},l^{\prime\prime},m^{\prime\prime}%
}\left\langle \psi_{n,l,m}\left\vert \hat{z}\right\vert \psi_{n^{\prime\prime
},l^{\prime\prime},m^{\prime\prime}}\right\rangle \nonumber\\
&  \times\left\langle \psi_{n^{\prime},l^{\prime},m^{\prime}}\left\vert
\hat{z}\right\vert \psi_{0}\right\rangle \nonumber\\
&  \times\left\langle 0_{ph}\left\vert \left\langle \psi_{0}\left\vert
e^{it\hat{H}^{\prime}}\right\vert \psi_{n,l,m}\right\rangle \right.  \right.
\nonumber\\
&  \times\left.  \left.  \left\langle \psi_{n^{\prime\prime},l^{\prime\prime
},m^{\prime\prime}}\left\vert e^{-it\hat{H}^{\prime}}\right\vert
\psi_{n^{\prime},l^{\prime},m^{\prime}}\right\rangle \right\vert
0_{ph}\right\rangle . \label{fzz3}%
\end{align}
where $\left\vert \psi_{n,l,m}\right\rangle $ are the polaron states as
obtained within the strong coupling ansatz in Ref. \cite{PRB2014}. The
transformed Hamiltonian $\hat{H}^{\prime}$ of the electron-phonon system after
the strong coupling unitary transformation \cite{PRB2014} takes the form%
\begin{equation}
\hat{H}^{\prime}=\hat{H}_{0}^{\prime}+\hat{W} \label{HT2}%
\end{equation}
with the terms%
\begin{align}
\hat{H}_{0}^{\prime}  &  =\frac{\mathbf{\hat{p}}^{2}}{2}+\sum_{\mathbf{q}%
}\left\vert f_{\mathbf{q}}\right\vert ^{2}+V_{a}\left(  \mathbf{\hat{r}%
}\right)  +\sum_{\mathbf{q}}\left(  \hat{b}_{\mathbf{q}}^{+}\hat
{b}_{\mathbf{q}}+\frac{1}{2}\right)  ,\label{H0}\\
\hat{W}  &  =\sum_{\mathbf{q}}\left(  \hat{w}_{\mathbf{q}}\hat{b}_{\mathbf{q}%
}+\hat{w}_{\mathbf{q}}^{\ast}\hat{b}_{\mathbf{q}}^{+}\right)  . \label{W}%
\end{align}
Here, $w_{\mathbf{q}}$ are the amplitudes of the renormalized electron-phonon
interaction%
\begin{equation}
\hat{w}_{\mathbf{q}}=\frac{\sqrt{2\sqrt{2}\pi\alpha}}{q\sqrt{V}}\left(
e^{i\mathbf{q\cdot\hat{r}}}-\rho_{\mathbf{q},0}\right)  ,
\end{equation}
where $\rho_{\mathbf{q},0}$ is the expectation value of the operator
$e^{i\mathbf{q\cdot\hat{r}}}$ with the trial electron wave function
$\left\vert \psi_{0}\right\rangle $:%
\begin{equation}
\rho_{\mathbf{q},0}=\left\langle \psi_{0}\left\vert e^{i\mathbf{q\cdot\hat{r}%
}}\right\vert \psi_{0}\right\rangle ,
\end{equation}
and $V_{a}\left(  \mathbf{\hat{r}}\right)  $ is the self-consistent potential
energy for the electron,%
\begin{equation}
V_{a}\left(  \mathbf{\hat{r}}\right)  =-\sum_{\mathbf{q}}\frac{4\sqrt{2}%
\pi\alpha}{q^{2}V}\rho_{-\mathbf{q},0}e^{i\mathbf{q}\cdot\mathbf{\hat{r}}}.
\end{equation}
The eigenstates of the Hamiltonian $\hat{H}_{0}^{\prime}$ are the products of
the electron wave functions and those of the phonon vacuum $\left\vert
\psi_{n,l,m}\right\rangle \left\vert 0_{ph}\right\rangle $. The dipole-dipole
correlation function $f_{zz}\left(  t\right)  $ given by (\ref{fzz4}) is
simplified within the adiabatic approximation \emph{for the ground state} and
using the selection rules for the dipole transition matrix elements and the
symmetry properties of the polaron Hamiltonian, as in Ref. \cite{PRB2014}. The
correlation function, using the interaction representation, takes the form,
\begin{align}
f_{zz}\left(  t\right)   &  =\sum_{n^{\prime},n}\left\langle \psi
_{0}\left\vert \hat{z}\right\vert \psi_{n,1,0}\right\rangle \left\langle
\psi_{n^{\prime},1,0}\left\vert \hat{z}\right\vert \psi_{0}\right\rangle
e^{-i\Omega_{n,0}t}\nonumber\\
&  \times\left\langle \psi_{n,1,0}\left\vert \left\langle 0_{ph}\left\vert
\mathcal{U}\left(  t\right)  \right\vert 0_{ph}\right\rangle \right\vert
\psi_{n^{\prime},1,0}\right\rangle , \label{fzz8}%
\end{align}
with the Franck-Condon transition frequency,
\begin{equation}
\Omega_{n,0}\equiv\varepsilon_{n,1}-\varepsilon_{1,0},
\end{equation}
and the evolution operator%
\begin{equation}
\mathcal{U}\left(  t\right)  =\mathrm{T}\exp\left[  -i\int_{0}^{t}ds\hat
{W}\left(  s\right)  \right]  ,
\end{equation}
where $\mathrm{T}$ is the time-ordering symbol, and $\hat{W}\left(  s\right)
$ is the interaction Hamiltonian in the interaction representation,%
\begin{equation}
\hat{W}\left(  s\right)  =e^{i\hat{H}^{\prime}s}\hat{W}e^{-i\hat{H}^{\prime}%
s}.
\end{equation}

As found in early works on the strong-coupling Fr\"{o}hlich polaron (see, for
review, Refs. \cite{Pekar1954,Allcock}), the energy differences between
different excited FC states for a strong coupling polaron are much smaller
than the energy difference between the ground and lowest excited FC state.
Therefore we keep here the adiabatic approximation for the ground state and,
consequently, for the transition between the ground and excited states. On the
contrary, the adiabatic approximation for the transitions between
\emph{different excited states} is \emph{not} applied in (\ref{fzz8}), as
distinct from the calculation in Ref. \cite{PRB2014}.

The matrix elements for the dipole transitions from the ground state to other
excited states than $\left\vert \psi_{1,1,0}\right\rangle $ (i. e.,
$\left\langle \psi_{0}\left\vert z\right\vert \psi_{n,1,0}\right\rangle $ with
$n\neq1$) have small relative oscillator strengths with respect to
$\left\langle \psi_{0}\left\vert z\right\vert \psi_{1,1,0}\right\rangle $ (of
order $\sim10^{-2}$). Therefore further on we consider the next-to-leading
order nonadiabatic corrections for the contribution to (\ref{fzz8}) with
$n=n^{\prime}=1$ and the adiabatic expression for the contribution with the
other terms. In this approximation, (\ref{fzz8}) takes the form%
\begin{align}
f_{zz}\left(  t\right)   &  =\sum_{n}\left\vert \left\langle \psi
_{0}\left\vert \hat{z}\right\vert \psi_{n,1,0}\right\rangle \right\vert
^{2}e^{-i\Omega_{n,0}t}\nonumber\\
&  \times\left\langle \psi_{n,1,0}\left\vert \left\langle 0_{ph}\left\vert
\mathcal{U}\left(  t\right)  \right\vert 0_{ph}\right\rangle \right\vert
\psi_{n,1,0}\right\rangle . \label{fzz9}%
\end{align}
The exact averaging over the phonon variables is performed by the
disentangling of the evolution operator (in analogy with \cite{Feynman1951}).
As a result, we obtain the formula%
\begin{align}
f_{zz}\left(  t\right)   &  =\sum_{n}\left\vert \left\langle \psi
_{0}\left\vert z\right\vert \psi_{n,1,0}\right\rangle \right\vert
^{2}e^{-i\Omega_{n,0}t}\nonumber\\
&  \times\left\langle \psi_{n,1,0}\left\vert \mathrm{T}_{e}\exp\left(
\hat{\Phi}\right)  \right\vert \psi_{n,1,0}\right\rangle \label{B}%
\end{align}
with the \textquotedblleft influence phase\textquotedblright\ (assuming
$\hbar=1$ and $\omega_{0}=1$)%
\begin{equation}
\hat{\Phi}=-\int_{0}^{t}ds\int_{0}^{s}ds^{\prime}e^{-i\left(  s-s^{\prime
}\right)  }\sum_{\mathbf{q}}\hat{w}_{\mathbf{q}}\left(  s\right)  \hat
{w}_{\mathbf{q}}^{+}\left(  s^{\prime}\right)  ,
\end{equation}
and $\mathrm{T}_{e}$ the time-ordering symbol with respect to the electron
degrees of freedom. The correlation function (\ref{B}) is the basis expression
for the further treatment.

The next approximation is the restriction to the leading-order semi-invariant
expansion:%
\begin{align}
&  \left\langle \psi_{n,1,0}\left\vert \mathrm{T}_{e}\exp\left(  \hat{\Phi
}\right)  \right\vert \psi_{n,1,0}\right\rangle \nonumber\\
&  \approx\exp\left\langle \psi_{n,1,0}\left\vert \mathrm{T}_{e}\left(
\hat{\Phi}\right)  \right\vert \psi_{n,1,0}\right\rangle .
\end{align}
As shown in Ref. \cite{PRB2014}, this approximation accounts of the static
Jahn-Teller effect, and it works well, because the dynamic Jahn-Teller effect
appears to be very small. The influence phase is invariant under spatial
rotations so that
\begin{align}
&  \left\langle \psi_{n,1,0}\left\vert \mathrm{T}_{e}\left(  \hat{\Phi
}\right)  \right\vert \psi_{n,1,0}\right\rangle \nonumber\\
&  =\left\langle \psi_{n,1,1}\left\vert \mathrm{T}_{e}\left(  \hat{\Phi
}\right)  \right\vert \psi_{n,1,1}\right\rangle \nonumber\\
&  =\left\langle \psi_{n,1,-1}\left\vert \mathrm{T}_{e}\left(  \hat{\Phi
}\right)  \right\vert \psi_{n,1,-1}\right\rangle .
\end{align}
Hence the correlation function (\ref{B}) can be simplified to%
\begin{align}
f_{zz}\left(  t\right)   &  =\sum_{n}\left\vert \left\langle \psi
_{0}\left\vert \hat{z}\right\vert \psi_{n,1,0}\right\rangle \right\vert
^{2}\nonumber\\
&  \times\exp\left(  -i\Omega_{n,0}t%
\genfrac{}{}{0pt}{0}{{}}{{}}%
\right. \nonumber\\
&  -\frac{1}{3}\sum_{\mathbf{q}}\sum_{n^{\prime},l^{\prime},m^{\prime}%
,m}\left\vert \left\langle \psi_{n,1,m}\left\vert \hat{w}_{\mathbf{q}%
}\right\vert \psi_{n^{\prime},l^{\prime},m^{\prime}}\right\rangle \right\vert
^{2}\nonumber\\
&  \left.  \times\frac{1-i\omega_{n^{\prime},l^{\prime};n,1}t-e^{-i\omega
_{n^{\prime},l^{\prime};n,1}t}}{\omega_{n^{\prime},l^{\prime};n,1}^{2}%
}\right)  . \label{fzz1}%
\end{align}
with the notation%
\begin{equation}
\omega_{n^{\prime},l^{\prime};n,1}\equiv1+\varepsilon_{n^{\prime},l^{\prime}%
}-\varepsilon_{n,1}. \label{w}%
\end{equation}

In our previous treatments of the strong coupling polaron optical
conductivity, we neglected the matrix elements for $\hat{w}_{\mathbf{q}}$
between the electron energy levels with different energies. Physically, this
approximation corresponds to the adiabatic approximation. Here, we go beyond
this approximation, taking into account the transitions between different
excited states but still assuming (as before) that the adiabatic approximation
holds for the transitions between the ground and excited states. In other
words, we keep in the sum $\sum_{n^{\prime},l^{\prime},m^{\prime},m}$ in
(\ref{fzz1}) only the \emph{excited} states.

Introducing parameters related to the extension of the Huang-Rhys parameter
used in Ref. \cite{PRB2014}:%
\begin{equation}
S_{n^{\prime},l;n,1}\equiv\frac{1}{3\omega_{n^{\prime},l;n,1}^{2}}%
\sum_{\mathbf{q}}\sum_{m^{\prime},m}\left\vert \left\langle \psi
_{n,1,m}\left\vert \hat{w}_{\mathbf{q}}\right\vert \psi_{n^{\prime
},l,m^{\prime}}\right\rangle \right\vert ^{2}, \label{Snl}%
\end{equation}
the correlation function is rewritten as follows:%
\begin{align}
&  f_{zz}\left(  t\right) \nonumber\\
&  =\sum_{n}\left\vert \left\langle \psi_{0}\left\vert z\right\vert
\psi_{n,1,0}\right\rangle \right\vert ^{2}\exp\left[  -i\Omega_{n,0}%
t-\sum_{n^{\prime},l}S_{n^{\prime},l;n,1}\right. \nonumber\\
&  \left.
\genfrac{}{}{0pt}{0}{{}}{{}}%
\times\left(  1-i\omega_{n^{\prime},l;n,1}t-e^{-i\omega_{n^{\prime},l;n,1}%
t}\right)  \right]  . \label{fzz2}%
\end{align}
The states $\left\vert \psi_{n^{\prime},l,m^{\prime}}\right\rangle $ can be
subdivided to two groups: (1) the states $\left\vert \psi_{1,1,m^{\prime}%
}\right\rangle $ with the energy level $\varepsilon_{1,1}$, (2) the higher
energy states with $\left(  n^{\prime},l\right)  \neq\left(  1,1\right)  $.
The first group of states were already taken into account in our previous
treatments and in Ref. \cite{PRB2014}. Taking into account the second group of
states provides the step beyond the adiabatic approximation -- this is the
focus of the present treatment. We denote the parameters corresponding to the
adiabatic approximation by%
\begin{equation}
S_{n}\equiv S_{n,1;n,1}\equiv\frac{1}{3}\sum_{\mathbf{q}}\sum_{m^{\prime}%
,m}\left\vert \left\langle \psi_{n,1,m}\left\vert \hat{w}_{\mathbf{q}%
}\right\vert \psi_{n,1,m^{\prime}}\right\rangle \right\vert ^{2}. \label{S0}%
\end{equation}
Correspondingly, the correlation function (\ref{fzz2}) takes the form%
\begin{align}
f_{zz}\left(  t\right)   &  =\sum_{n}\left\vert \left\langle \psi
_{0}\left\vert z\right\vert \psi_{n,1,0}\right\rangle \right\vert
^{2}\nonumber\\
&  \times\exp\left[  -i\Omega_{n,0}t-S_{n}\left(  1-it-e^{-it}\right)
\genfrac{}{}{0pt}{0}{{}}{{}}%
\right. \nonumber\\
&  -\sum\nolimits_{\left(  n^{\prime},l\right)  \neq\left(  n,1\right)
}S_{n^{\prime},l;n,1}\nonumber\\
&  \left.  \times\left(  1-i\omega_{n^{\prime},l;n,1}t-e^{-i\omega_{n^{\prime
},l;n,1}t}\right)  \right]  . \label{fzz4}%
\end{align}
Following Refs. \cite{PRL2006,PRB2014}, the factor $\left(  1-it-e^{-it}%
\right)  $ is expanded in powers of $t$:%
\begin{equation}
1-it-e^{-it}=\frac{1}{2}t^{2}+O\left(  t^{3}\right)  . \label{exp}%
\end{equation}
This approximation is appropriate in the strong coupling regime, where the
phonon frequency is small with respect to other frequencies, such as the
Franck-Condon frequency $\Omega_{1,0}$, which increases as $\Omega
_{1,0}\propto\alpha^{2}$ at large $\alpha$. In the case when non-adiabatic
terms are non taken into account, the expansion (\ref{exp}) provides an
envelope of the optical conductivity spectrum obtained in
\cite{PRL2006,PRB2014}. The other factor, $\left(  1-i\omega_{n^{\prime
},l;n,1}t-e^{-i\omega_{n^{\prime},l;n,1}t}\right)  $, should not be expanded
in the same way, because the frequencies $\omega_{n^{\prime},l;n,1}$ $\left(
n^{\prime},l\right)  \neq\left(  1,1\right)  $ also increase in the strong
coupling limit as $\alpha^{2}$. Therefore we keep this factor as is, without
expansion. As a result, in the strong coupling regime the correlation function
$f_{zz}\left(  t\right)  $ is:%
\begin{align}
f_{zz}\left(  t\right)   &  =\sum_{n}\left\vert \left\langle \psi
_{0}\left\vert z\right\vert \psi_{n,1,0}\right\rangle \right\vert
^{2}\nonumber\\
&  \times\exp\left(  -\delta S_{n}-i\tilde{\Omega}_{n,0}t-\frac{1}{2}%
S_{n}t^{2}\right. \nonumber\\
&  \left.  +\sum_{\left(  n^{\prime},l\right)  \neq\left(  n,1\right)
}S_{n^{\prime},l;n,1}e^{-i\omega_{n^{\prime},l;n,1}t}\right)  . \label{fzz5}%
\end{align}
with the parameters:%
\begin{align}
\delta S_{n}  &  \equiv\sum_{\left(  n^{\prime},l\right)  \neq\left(
1,1\right)  }S_{n^{\prime},l;n,1},\label{p1}\\
\delta\Omega_{n}  &  \equiv\sum_{\left(  n^{\prime},l\right)  \neq\left(
1,1\right)  }S_{n^{\prime},l;n,1}\omega_{n^{\prime},l;n,1},\label{p2}\\
\tilde{\Omega}_{n,0}  &  \equiv\Omega_{n,0}-\delta\Omega_{n}. \label{p3}%
\end{align}
The parameter $\delta S_{n}$ plays a role of the Debye-Waller factor and
ensures the fulfilment of the $f$-sum rule for the optical conductivity. The
parameter $\delta\Omega_{n}$ is the shift of the Franck-Condon frequency to a
lower value due to phonon-assisted transitions to higher energy states. The
exponent can be expanded, yielding a description in terms of multiphonon
processes:%
\begin{align}
&  \exp\left(  \sum_{\left(  n^{\prime},l\right)  \neq\left(  n,1\right)
}S_{n^{\prime},l;n,1}e^{-i\omega_{n^{\prime},l;n,1}t}\right) \nonumber\\
&  =\sum_{\left\{  p_{n^{\prime},l}\geq0\right\}  }\left(  \prod_{\left(
n^{\prime},l\right)  \neq\left(  n,1\right)  }\frac{S_{n^{\prime}%
,l;n,1}^{p_{n^{\prime},l;n,1}}}{p_{n^{\prime},l;n,1}!}\right) \nonumber\\
&  \times e^{-i\sum_{n^{\prime},l}p_{n^{\prime},l;n,1}\omega_{n^{\prime
},l;n,1}t}, \label{expan}%
\end{align}
where the sum $\sum_{\left\{  p_{n^{\prime},l}\right\}  }$ is performed over
all combinations $\left\{  p_{n^{\prime},l}\geq0\right\}  $.

With the expansion (\ref{expan}), the polaron optical conductivity takes the
form:%
\begin{align}
&  \operatorname{Re}\sigma\left(  \Omega\right) \nonumber\\
&  =\Omega\sum_{n}\left\vert \left\langle \psi_{0}\left\vert z\right\vert
\psi_{n,1,0}\right\rangle \right\vert ^{2}e^{-\delta S_{n}}\sqrt{\frac{\pi
}{2S_{n}}}\nonumber\\
&  \times\sum_{\left\{  p_{n^{\prime},l;n,1}\geq0\right\}  }\left(
\prod_{\left(  n^{\prime},l\right)  \neq\left(  n,1\right)  }\frac
{S_{n^{\prime},l;n,1}^{p_{n^{\prime},l;n,1}}}{p_{n^{\prime},l;n,1}!}\right)
\nonumber\\
&  \times\exp\left[  -\frac{\left(  \tilde{\Omega}_{n,0}+\sum_{n^{\prime}%
,l}p_{n^{\prime},l;n,1}\omega_{n^{\prime},l;n,1}-\Omega\right)  ^{2}}{2S_{n}%
}\right]  . \label{ReS}%
\end{align}
In formula (\ref{ReS}), the term where all $p_{n^{\prime},l;n,1}=0$
corresponds to the adiabatic approximation and exactly reproduces the result
of Ref. \cite{PRB2014}. The other terms represent the \emph{non-adiabatic}
contributions to $\operatorname{Re}\sigma\left(  \Omega\right)  $, and are
correction terms to the previously found results.

\section{Results and discussions \label{sec:Results}}

The polaron optical conductivity derived in the above section is in line with
the physical understanding of the underlying processes for the polaron optical
response, achieved in early works \cite{KED1969,DSG} and summarized in Ref.
\cite{Devreese72}. It is based on the concept of the polaron excitations of
three types:

\begin{itemize}
\item Relaxed Excited States (RES) \cite{KED1969} for which the lattice
polarization is adapted to the electronic distribution;

\item Franck-Condon states (FC) where the lattice polarization is
\textquotedblleft frozen\textquotedblright,\ adapted to the polaron ground state;

\item Scattering states characterized by the presence of real phonons along
with the polaron.
\end{itemize}

The polaron RES can be formed when the electron-phonon coupling is strong
enough, for $\alpha\gtrapprox4.5$. At weak coupling, the polaron optical
response at zero temperature is due to transitions from the polaron ground
state to scattering states. In other words, the optical absorption spectrum of
a weak-coupling polaron is determined by the absorption of radiation energy,
which is re-emitted in the form of LO phonons. At stronger couplings, the
concept of the polaron relaxed excited states first introduced in Ref.
\cite{KED1969} becomes of key importance. In the range of sufficiently large
$\alpha$ when the polaron RES are formed, the absorption of light by a polaron
occurs through transitions from the ground state to RES which can be
accompanied by the emission of different numbers $n\geq0$ of free phonons.
These transitions contribute to the shape of a multiphonon optical absorption
spectrum. At very large coupling, lattice relaxation processes become to slow
and the Franck-Condon states determine the optical response.

We analyze polaron optical conductivity spectra both with the memory function
formalism and with the strong-coupling expansion, and compare these to the
DQMC numerical data \cite{M2003}. Within the framework of formalisms based on
the memory function (MF), we compare the following theories:

\begin{itemize}
\item The original DSG method of Ref. \cite{DSG}, where the expectation value
in \ref{xi4} is calculated with respect to a Gaussian trial action. This will
be denoted by \emph{MF-1} in the figures.

\item The extended MF formalism of \cite{PRL2006}, where an ad-hoc broadening
with a strength determined from sum rules is included in (\ref{4}). This will
be denoted by \emph{MF-2}.

\item The current non-quadratic MF formalism, based on the extension of the
Jensen-Feynman inequality introduced in this paper, denoted by \emph{MF-new}.
\end{itemize}

\noindent Among the strong-coupling expansions (SCE), we distinguish:

\begin{itemize}
\item The strong-coupling result in the adiabatic approximation, as obtained
in Ref. \cite{PRL2006}. This will be denoted here by \emph{SCE-1}.

\item The adiabatic approximation of Ref. \cite{PRB2014}, which uses more
accurate trial polaron states.\ This will be denoted by \emph{SCE-2}.

\item The current non-adiabatic strong coupling expansion, denoted by
\emph{SCE-new}.
\end{itemize}

%

\begin{figure}
[h]
\begin{center}
\includegraphics[
height=5.8721in,
width=2.6195in
]%
{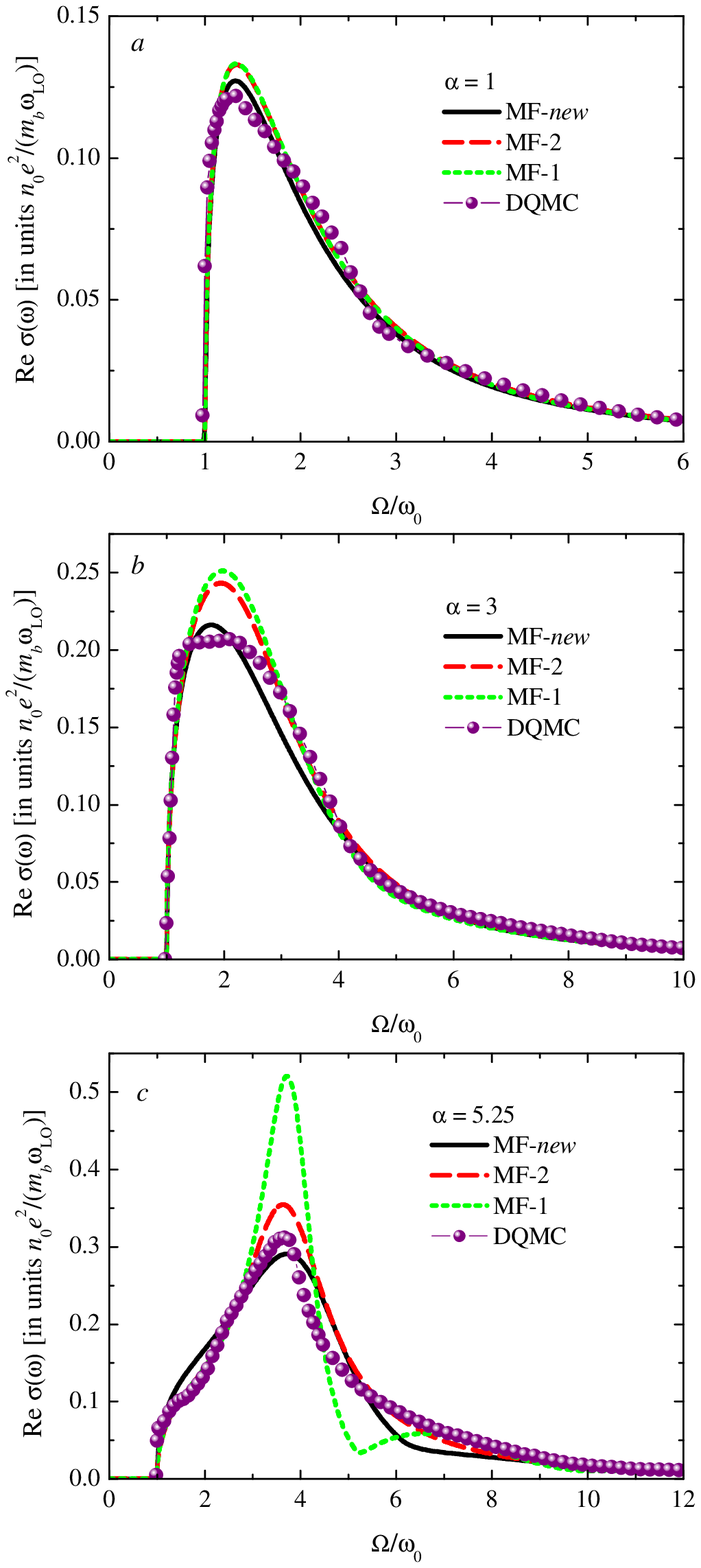}%
\caption{Polaron optical conductivity calculated for $\alpha=1$ (\emph{a}),
$\alpha=3$ (\emph{b}) and $\alpha=5.25$ (\emph{c}) within the present
non-quadratic MF formalism (denoted in the figure as MF-\emph{new}), compared
with the polaron optical conductivity calculated within the extended
memory-function formalism (MF-2) of Ref. \cite{PRL2006}, the results of the
memory-function approach using the Feynman parabolic trial action \cite{DSG}
(MF-1), and the diagrammatic quantum Monte Carlo (DQMC) \cite{M2003,PRL2006}.}%
\label{Figure1}%
\end{center}
\end{figure}

The subsequent figures show the results for increasing $\alpha$. In Figure
\ref{Figure1}, the optical conductivity is shown for small coupling,
$\alpha=1,$and for $\alpha=3,5.25$ which correspond to the dynamic regime
where the RES starts to play a role. In this regime, analytic solutions are
provided by the various memory function formalisms listed above, and we
compare them to DQMC numeric data \cite{M2003}. At weak coupling
($\alpha=1\,,$ panel (\emph{a})$)$, all the approaches based on the memory
function give results in agreement with DQMC. For $\alpha=3$ (panel
(\emph{b})), the current method gives a better fit to the DQMC result that the
other two methods. For a stronger coupling, $\alpha=5.25$ (panel (\emph{c}))
the MF-2 approach substantially improves the original result MF-1, but the
optical conductivity spectrum calculated within the new non-quadratic MF
formalism lies closer to the DQMC data than either of the other two.

\begin{figure}
[h]
\begin{center}
\includegraphics[
height=3.9297in,
width=2.6022in
]%
{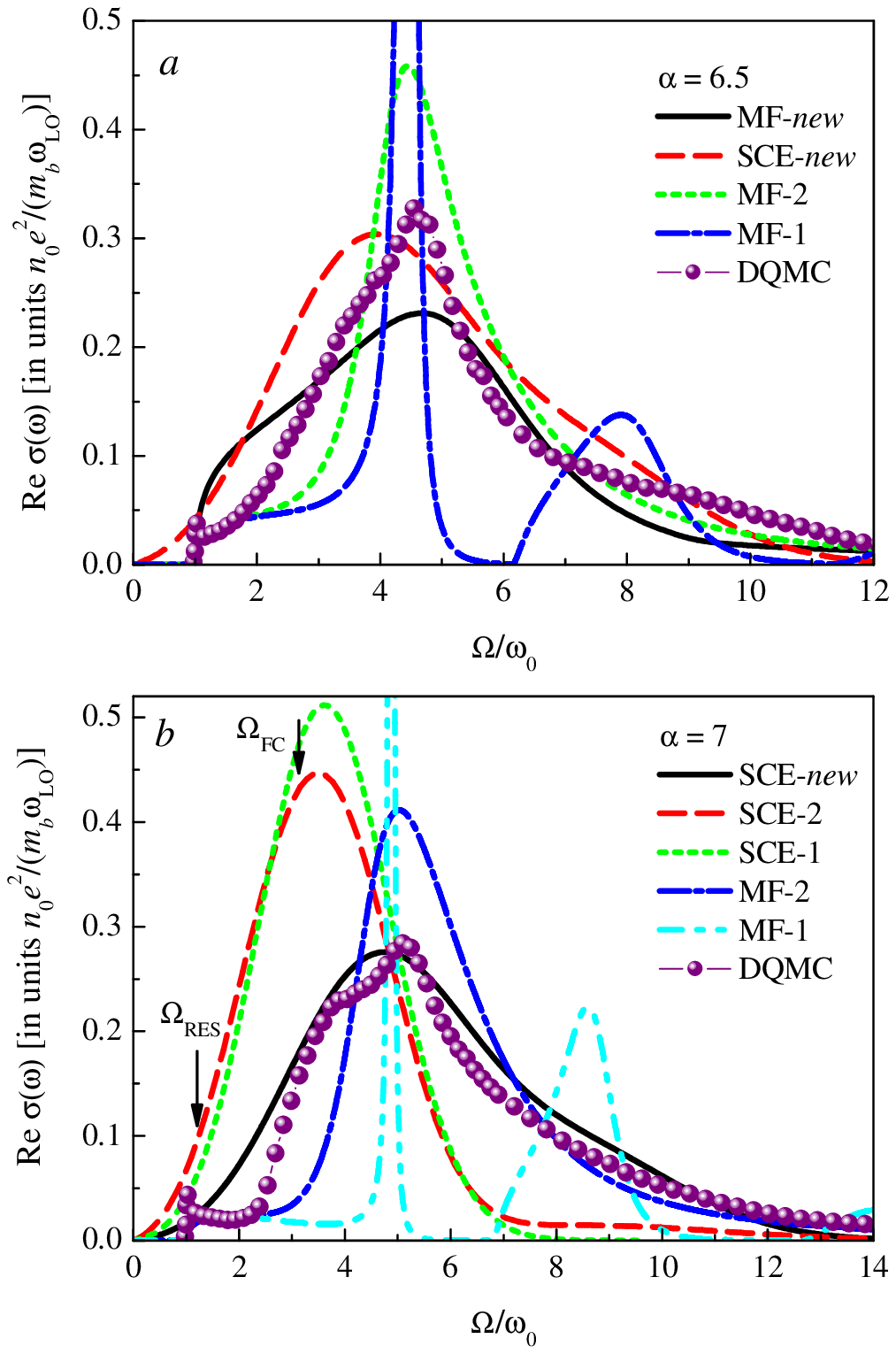}%
\caption{Polaron optical conductivity calculated for $\alpha=6.5$ (\emph{a})
and $\alpha=7$ (\emph{b}) using different analytic approaches: the
non-quadratic MF formalism (MF-\emph{new}), the extended memory-function
formalism of Ref. \cite{PRL2006} (MF-2), the memory-function approach with the
Feynman parabolic trial action \cite{DSG} (MF-1), the non-adiabatic
strong-coupling expansion (denoted at the figure as SCE-\emph{new}), the
adiabatic strong-coupling expansions of Refs. \cite{PRL2006,PRB2014} (SCE-1
and SCE-2). The results are compared to DQMC data of Refs.
\cite{M2003,PRL2006}.}%
\label{Figure2}%
\end{center}
\end{figure}

Fig. \ref{Figure2} demonstrates the behavior of the polaron optical
conductivity spectra in the intermediate coupling regime, for $\alpha=6.5$ and
$\alpha=7$. In this regime, the existing memory function approaches
(\emph{MF-1},\emph{MF-2}) as well as the existing strong coupling expansions
(\emph{SCE-1},\emph{SCE-2}) do not provide satisfactory results. The new
memory function approach and the new strong coupling expansion are in much
better agreement with the DQMC data.%

This range of coupling parameters is where one would want to cross over from
using a memory function based approach to a strong coupling expansion. Whereas
the existing methods do not allow to bridge this gap at intermediate coupling,
the extensions that we have proposed here are suited to implement such a
cross-over. The present memory-function approach with the non-parabolic trial
action leads to a relatively small extension of the range of $\alpha$ where
the polaron optical conductivity compares well with the DQMC data, namely from
$\alpha\approx4.5$ to $\alpha\approx6.5$. For $\alpha\lessapprox6.5$,
the$\,$memory-function approach with the non-parabolic trial action provides a
better agreement with DQMC than all other known approximations. Remarkably,
the optical conductivity spectra as given by the non-quadratic MF formalism
and the non-adiabatic SCE are both in better agreement with the Monte Carlo
data than any of the preceding analytical methods. For $\alpha=6.5$, the
polaron optical conductivity calculated within non-quadratic MF formalism and
the non-adiabatic SCE lie rather close to each other. We can conclude
therefore that the ranges of validity of those two approximations overlap,
despite the fact that these approximations are based on different assumptions.

The maximum of the optical conductivity spectrum provided by the non-quadratic
MF formalism for $\alpha=6.5$ is positioned at slightly lower frequency than
that for the maximum of the optical conductivity obtained in the strong
coupling approximation with non-adiabatic corrections. They lie remarkably
close to two features of the DQMC optical conductivity spectrum: the
higher-frequency peak, which is the maximum of the spectrum, and the
lower-frequency shoulder. The similar comparative behavior of the
memory-function and strong coupling results was noticed in Ref. \cite{PRL2006}%
, where it was suggested that these two features in the DQMC spectra can
correspond physically to the dynamic (RES) and the Franck-Condon
contributions. The present results are in line with that physical picture.%

\begin{figure}
[h]
\begin{center}
\includegraphics[
height=5.8444in,
width=2.5927in
]%
{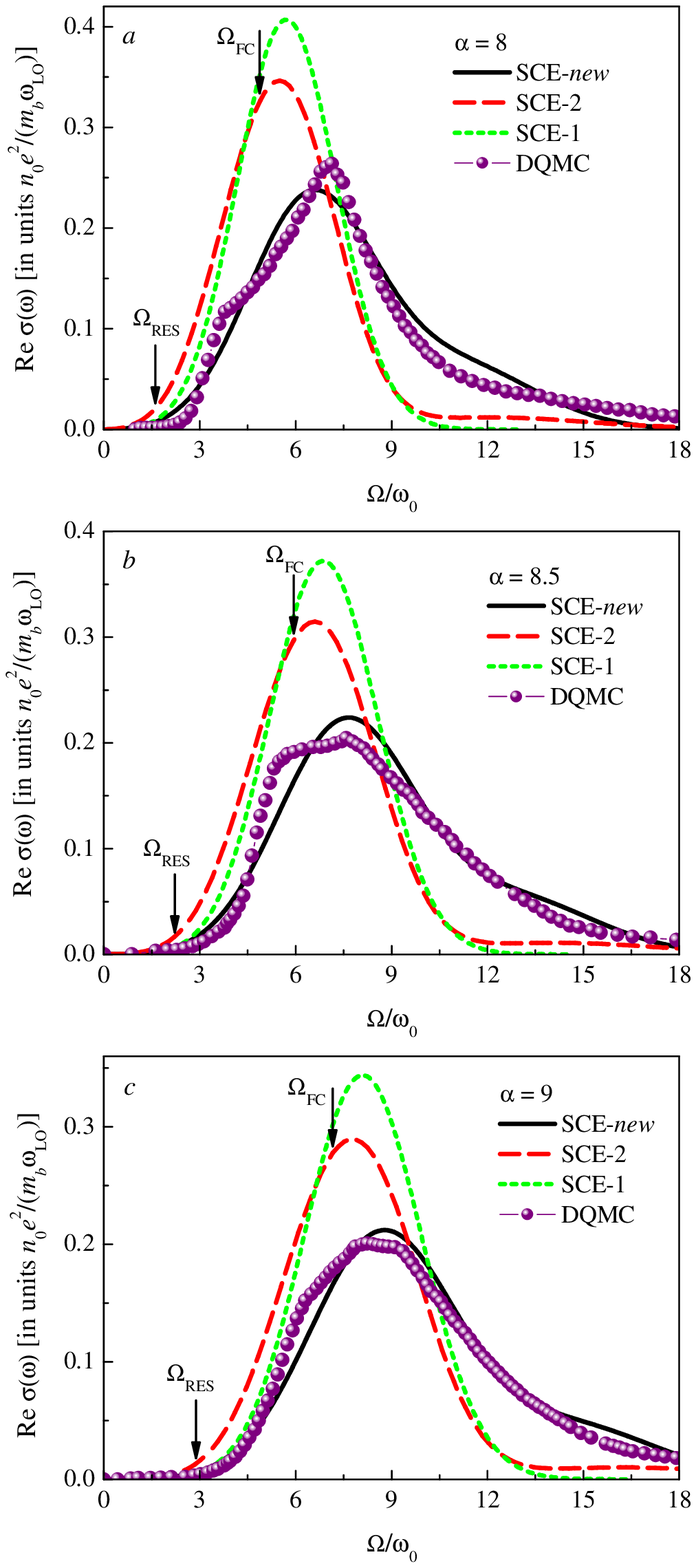}%
\caption{Polaron optical conductivity calculated for $\alpha=8$ (\emph{a}),
$\alpha=8.5$ (\emph{b}) and $\alpha=9$ (\emph{c}) within several analytic
strong coupling approaches and compared to DQMC data of Refs.
\cite{M2003,PRL2006}. The notations are the same as in Fig. 2.}%
\label{Figure3}%
\end{center}
\end{figure}

\begin{figure}
[h]
\begin{center}
\includegraphics[
height=3.87in,
width=2.6377in
]%
{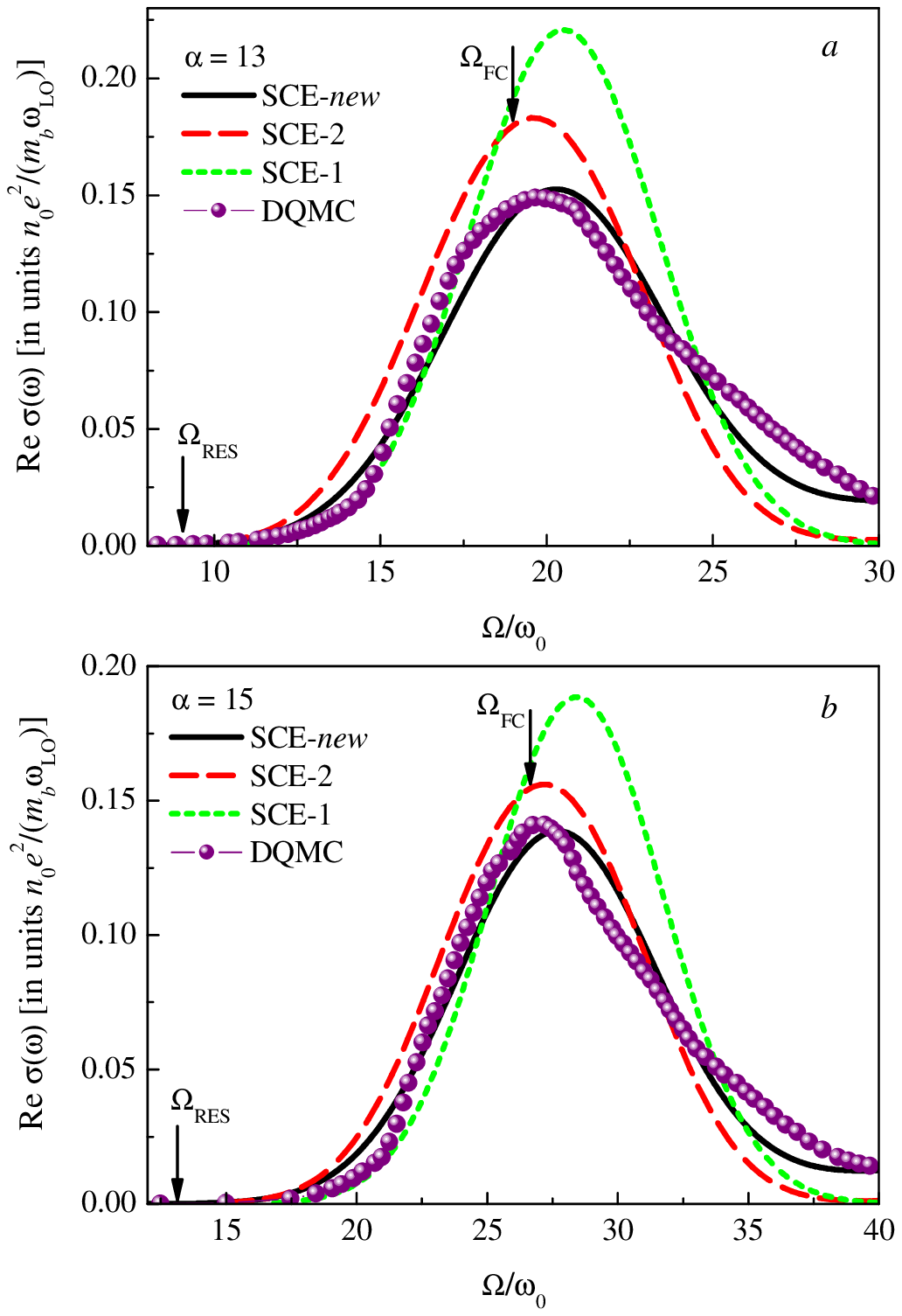}%
\caption{Polaron optical conductivity in the extremely strong coupling regime,
for $\alpha=13$ (\emph{a}) and $\alpha=15$ (\emph{b}). The notations are the
same as in Fig. 2.}%
\label{Figure4}%
\end{center}
\end{figure}

In Fig. \ref{Figure2}~(\emph{b}), the arrows indicate the FC transition
frequency for the transition to the first excited FC state $\Omega_{1,0}%
\equiv\Omega_{\mathrm{FC}}$ and the RES transition frequency $\Omega
_{\mathrm{RES}}$ for a strong coupling polaron as calculated in Ref.
\cite{KED1969}. We can see that both the shape and the position of the maximum
of the optical conductivity band obtained within the adiabatic approximation
in Refs. \cite{PRL2006,PRB2014} are rather far from those for the DQMC data.
Taking into account non-adiabatic transitions drastically improves the
agreement of the strong coupling approximation with DQMC, even for $\alpha=7$,
which, strictly speaking, is not yet the strong coupling regime. The value
$\alpha=7$ can be rather estimated as an intermediate coupling. However, even
at this intermediate coupling strength, the results of present approach lie
much closer to the DQMC data than those obtained within all other aforesaid
analytic methods. Also a substantial improvement of the agreement between the
strong coupling expansion and DQMC is clearly expressed in Fig. \ref{Figure3},
where the polaron optical conductivity spectra are shown for the strong
coupling regime for $\alpha=8$ to $\alpha=9$. For strong couplings, the
non-adiabatic SCE accurately reproduces both the peak position and the overall
shape of the DQMC spectra. Finally, we see that the results of the
non-adiabatic SCE remain accurate also in the extremely strong coupling
regime, as shown in Fig. \ref{Figure4}.%

\section{Conclusions \label{sec:Conclusions}}

In the present work, we have modified two basic analytic methods for the
polaron optical conductivity in order to extend their ranges of applicability
for the electron-phonon coupling constant in such a way that these ranges
overlap. The memory function formalism using a trial action for a model
two-particle system has been extended to work with non-quadratic interaction
potentials in the model system. This method combines the translation
invariance of the trial system, which is one of the main advantages of the
Feynman variational approach, with a more realistic interaction between the
electron and the fictitious particle. This extension leads to a substantial
improvement of the polaron optical conductivity for small and intermediate
coupling strengths with respect to the preceding known versions of the memory
function approach.

The other method is the strong-coupling expansion, and we have extended it
beyond the Franck-Condon adiabatic approximation by taking into account
non-adiabatic transitions between different excited polaron states. As a
result, the modified non-adiabatic strong-coupling expansion appears now to be
in good agreement with the numerical DQMC data in a wide range of $\alpha$
from intermediate coupling strength to the strong coupling limit. For the
intermediate coupling value $\alpha=6.5$, the two methods that we propose,
i.e. the non-quadratic MF formalism and the non-adiabatic SCE, result in
optical conductivity spectra which are remarkably close to each other and to
the DQMC results. Thus, both methods can be combined to provide all-coupling,
accurate analytic results for the polaron optical absorption.

For larger $\alpha$ the agreement between the results of the non-adiabatic SCE
and DQMC becomes gradually better. At very strong coupling, even the preceding
adiabatic SCE \cite{PRB2014} is already sufficiently good, so that the
improvement due to the non-adiabatic transitions, e.\ g., for $\alpha=15$, is
relatively small. However, for a slightly weaker coupling, e. g., for
$\alpha=9$, we can observe a drastically improved agreement with DQMC for the
present non-adiabatic SCE as compared to the adiabatic approximation. We can
conclude that at present, the strong coupling approximation taking into
account non-adiabatic contributions provides the best agreement with the DQMC
results for $\alpha\gtrapprox6.5$ with respect to all other known analytic
approaches for the polaron optical conductivity. We find that the
non-adiabatic transitions lead to a substantial change of the spectral shape
with respect to the optical conductivity derived within the adiabatic
approximation. The non-adiabatic effects are non-negligible in the whole range
of the coupling strength, at least for $\alpha\leq15$, available for DQMC.

In summary, extending the MF and SCE formalisms leads to an overlapping of the
areas of $\alpha$ where these two analytic methods are applicable. These
analytic methods have been verified, appearing to be in good agreement with
numeric DQMC data at all $\alpha$ available for DQMC. We therefore possess the
analytic description of the polaron optical response which embraces the whole
range of the coupling strength.

\begin{acknowledgments}
We thank A. S. Mishchenko for valuable discussions and the DQMC data for the
polaron optical conductivity, and V. Cataudella for the details of the EMFF
method. Discussions with F. Brosens and D. Sels are gratefully acknowledged.
This research has been supported by the Flemish Research Foundation (FWO-Vl),
project nrs. G.0115.12N, G.0119.12N, G.0122.12N, G.0429.15N, by the Scientific
Research Network of the Research Foundation-Flanders, WO.033.09N, and by the
Research Fund of the University of Antwerp.
\end{acknowledgments}


\begin{thebibliography}{99}                                                                                               %


\bibitem {Landau}L. D. Landau, \textit{Phys. Z. Sowjetunion} \textbf{3}, 664
(1933) [English translation in \textit{Collected Papers}, Gordon and Breach,
New York, 1965, pp. 67-68].

\bibitem {Devreese2009}A. S. Alexandrov and J. T. Devreese, \emph{Advances in
Polaron Physics} (Springer, 2009).

\bibitem {Hemolt}R. von Helmolt, J. Wecker, B. Holzapfel, L. Schultz, and K.
Samwer, Phys. Rev. Lett. \textbf{71}, 2331 (1993).

\bibitem {Sirringhaus}H. Sirringhaus \emph{et al}., Nature (London)
\textbf{401}, 685 (1999).

\bibitem {Holstein}T. Holstein, Ann. Phys. (N.Y.) \textbf{8}, 325 (1959).

\bibitem {Franchini1}M. Setvin, C. Franchini, X. Hao, M. Schmid, A. Janotti,
M. Kaltak, C. G. Van de Walle, G. Kresse, and U. Diebold, Phys. Rev. Lett.
\textbf{113}, 086402 (2014).

\bibitem {Franchini2015}X. Hao, Z. Wang, M. Schmid, U. Diebold, and C.
Franchini, Phys. Rev. B \textbf{91}, 085204 (2015).

\bibitem {BECpol2}J. Vlietinck, W. Casteels, K. Van Houcke, J. Tempere, J.
Ryckebusch, and J. T. Devreese, New J. Phys. \textbf{17}, 033023 (2015).

\bibitem {Meevasana}W. Meevasana, X. J. Zhou, B. Moritz, C.-C. Chen, R. H. He,
S.-I. Fujimori, D. H. Lu, S.-K. Mo, R. G. Moore, F. Baumberger, T. P.
Devereaux, D. van der Marel, N. Nagaosa, J. Zaanen and Z.-X. Shen, New Journal
of Physics \textbf{12}, 023004 (2010).

\bibitem {Mechelen2008}J. L. M. van Mechelen, D. van der Marel, C. Grimaldi,
A. B. Kuzmenko, N. P. Armitage, N. Reyren, H. Hagemann, and I. I. Mazin, Phys.
Rev. Lett. \textbf{100}, 226403 (2008).

\bibitem {PRB2010}J. T. Devreese, S. N. Klimin, J. L. M. van Mechelen, and D.
van der Marel, Phys. Rev. B \textbf{81}, 125119 (2010).

\bibitem {M2000}A. S. Mishchenko, N. V. Prokof'ev, A. Sakamoto, and
B.~V.~Svistunov, Phys. Rev. \textit{B} \textbf{62}, 6317 (2000).

\bibitem {M2003}A.\thinspace S. Mishchenko, N. Nagaosa, N.\thinspace V.
Prokof'ev, A. Sakamoto, and B.\thinspace V. Svistunov, Phys. Rev. Lett.
\textbf{91}, 236401 (2003).

\bibitem {Berciu}G. L. Goodvin, A. S. Mishchenko, and M. Berciu, Phys. Rev.
Lett. \textbf{107}, 076403 (2011).

\bibitem {KED1969}E.\ Kartheuser, R.\ Evrard, and J.\ Devreese Phys. Rev.
Lett. \textbf{22}, 94-97 (1969).

\bibitem {DSG}J. Devreese, J. De Sitter, and M. Goovaerts, Phys. Rev. B
\textbf{5}, 2367 (1972).

\bibitem {Devreese72}J. T. Devreese, in \emph{Polarons in Ionic Crystals and
Polar Semiconductors} (North-Holland, Amsterdam, 1972), pp. 83 -- 159.

\bibitem {PRL2006}G. De Filippis, V. Cataudella, A. S. Mishchenko, C. A.
Perroni, and J. T. Devreese, Phys. Rev. Lett. \textbf{96}, 136405 (2006).

\bibitem {GLF}V. L. Gurevich, I. G. Lang, and Yu. A. Firsov, Sov. Phys. Solid
State \textbf{4}, 918 (1962).

\bibitem {DHL1971}J. Devreese, W. Huybrechts, and L. Lemmens, Phys. Status
Solidi B \textbf{48}, 77 (1971).

\bibitem {PRB2014}S. N. Klimin and J. T. Devreese, Phys. Rev. B \textbf{89},
035201 (2014).

\bibitem {FHIP}R. P. Feynman, R. W. Hellwarth, C. K. Iddings, and P. M.
Platzman, Phys. Rev. \textbf{127}, 1004 (1962).

\bibitem {Feynman}R. P. Feynman, Phys. Rev. \textbf{97}, 660 (1955).

\bibitem {Dries1}D. Sels and F. Brosens, Phys. Rev. E \textbf{89}, 012124 (2014).

\bibitem {Dries2}D. Sels and F. Brosens, Phys.\ Rev. E \textbf{89}, 042110 (2014).

\bibitem {DS}D. Sels, \emph{arXiv:1605.04998} (2016).

\bibitem {Catau}V. Cataudella, G. De Filippis, and C.A. Perroni,
\textquotedblleft Single Polaron Properties in Different Electron-Phonon
Models\textquotedblright, in: \emph{Polarons in Advanced Materials}, ed. by A.
S. Alexandrov, Springer Series in Materials Science, Volume 103, 2007, pp. 149-189.

\bibitem {SSC}S. N. Klimin and J. T. Devreese, Solid State Communications
\textbf{151}, 144 (2011).

\bibitem {M75}S. J. Miyake, J. Phys. Soc. Japan \textbf{38}, 181 (1975).

\bibitem {Pekar1954}S. I. Pekar, \emph{Untersuchungen \"{u}ber die
Elektronentheorie der Kristalle} (Akademie-Verlag, Berlin, 1954).

\bibitem {Allcock}G. R. Allcock, in \emph{Polarons and Excitons}, edited by C.
G. Kuper and G. D. Whitfield (Oliver and Boyd, Edinburgh, 1963), pp. 45 -- 70.

\bibitem {MHolst}A. S. Mishchenko, N. Nagaosa, and N. Prokof'ev, Phys. Rev.
Lett. \textbf{113}, 166402 (2014).

\bibitem {PD1983}F. M. Peeters and J. T. Devreese, Phys. Rev. B \textbf{28},
6051 (1983).

\bibitem {Feynman1951}R. P. Feynman, Phys. Rev. \textbf{84}, 108 (1951).
\end{thebibliography}
\end{document}